# State Decoding in Multi-Stage Cryptography Protocols

Sindhu Chitikela

**Abstract.** This paper presents a practical method of quantum tomography for decoding the state of photons in a multistage cryptography protocol. This method works if the polarization angles are defined on a fixed plane, as is assumed in several quantum cryptography protocols. We show if there are $2^m$ polarization angles in a fixed plane, we need $m$ number of filters and $m^2$ number of photons through each filter.

## 1. Introduction

Photons are information carriers in quantum cryptography. In BB84 protocol [1], single photons represent a qubit whereas in multistage protocols [2]-[5], more than one photon could represent a bit. In the multistage protocol, the photon polarization is randomly changed by both Alice and Bob as in Figure 1 at first and then reversed so that Bob eventually receives the information Alice wanted to send. Clearly this system can be broken only if Eve can know the polarization state on each of the three links. This side-steps the constraint of BB84 that each bit be associated with a single photon. In two-stage and single stage versions [4], the initial set up to exchange the key is done by the three-stage protocol and once the key is shared then a single stage link with continually changing polarization angles according to a code that is included in the block of data sent [4] suffices for subsequent secure communication. The multistage protocol can be visualized in several kinds of state aware versions [5].

The multistage protocol was recently implemented [6],[7]. References [8] and [9] address the question of single photon generation and [10]-[12] address the question of implementation of quantum gates. References [13]-[15] consider general issues with practical quantum cryptography. Here we consider the question of state decoding in state aware quantum cryptography using a practical method of tomography that can help detect as eavesdropper who injects randomly polarized photons to compensate for the one’s she has siphoned off for state detection.

## 2. Tomography in multi-stage protocols

The state of the received photons is to be determined at the receiver. The equipment involved in the tomography process includes beamsplitters, half silvered mirrors and filters. The received stream of photons is sent through the beamsplitters or half silvered mirrors so that the stream is split to pass through different filters aligned at specific angles. Figure 2 is a schematic of the decoding process at the receiver. Figure 3 is a schematic of the quantum tomography process.

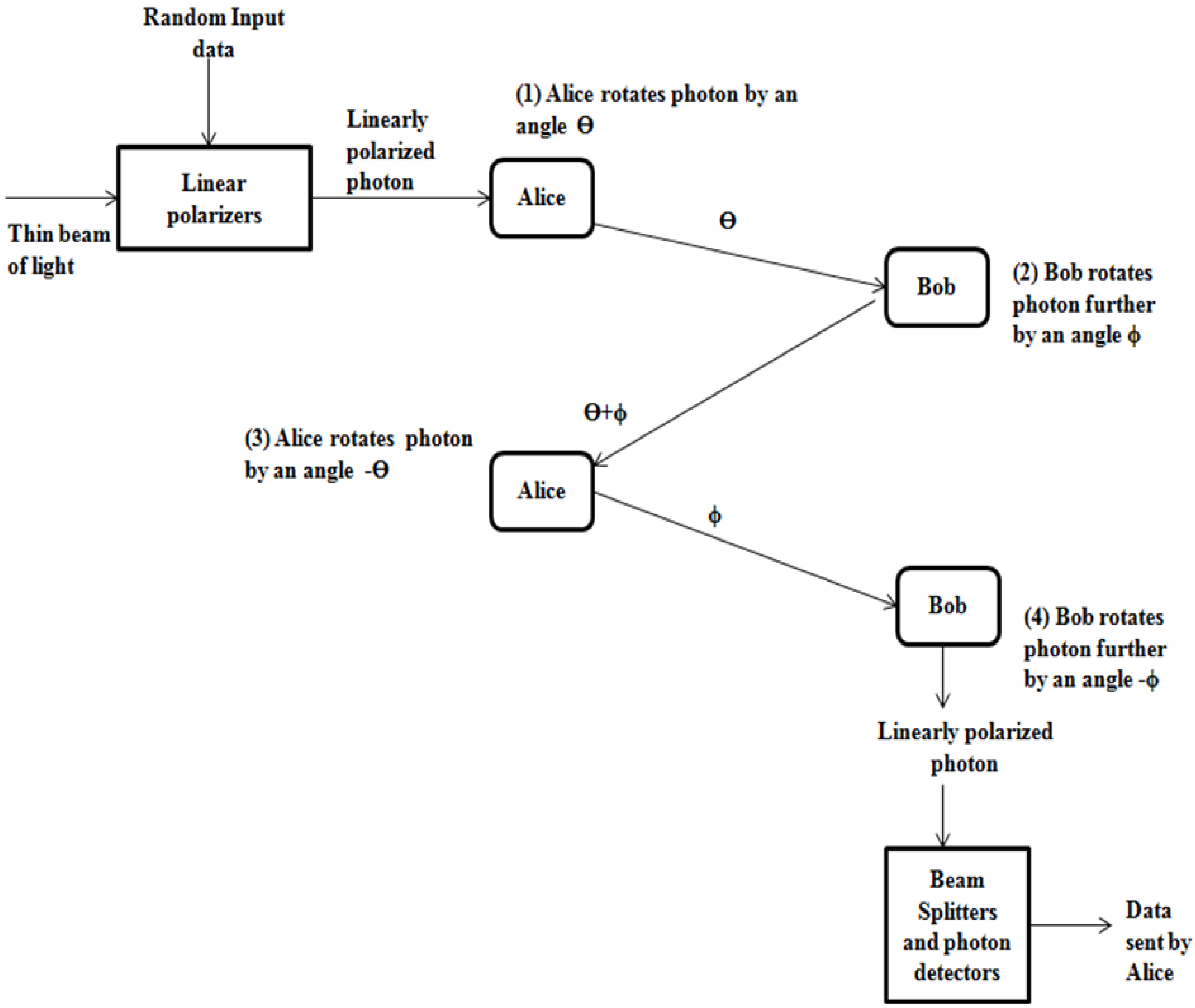

Figure 1 Implementation of the three-stage protocol

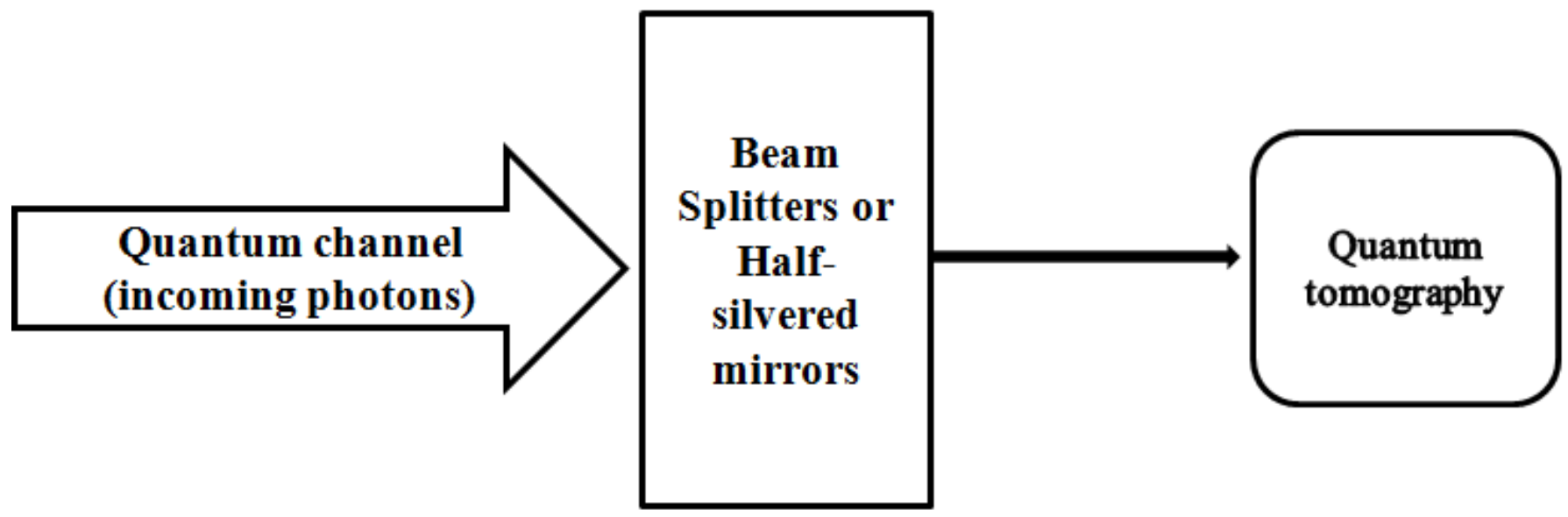

Figure 2 Information decoding process at receiver's site

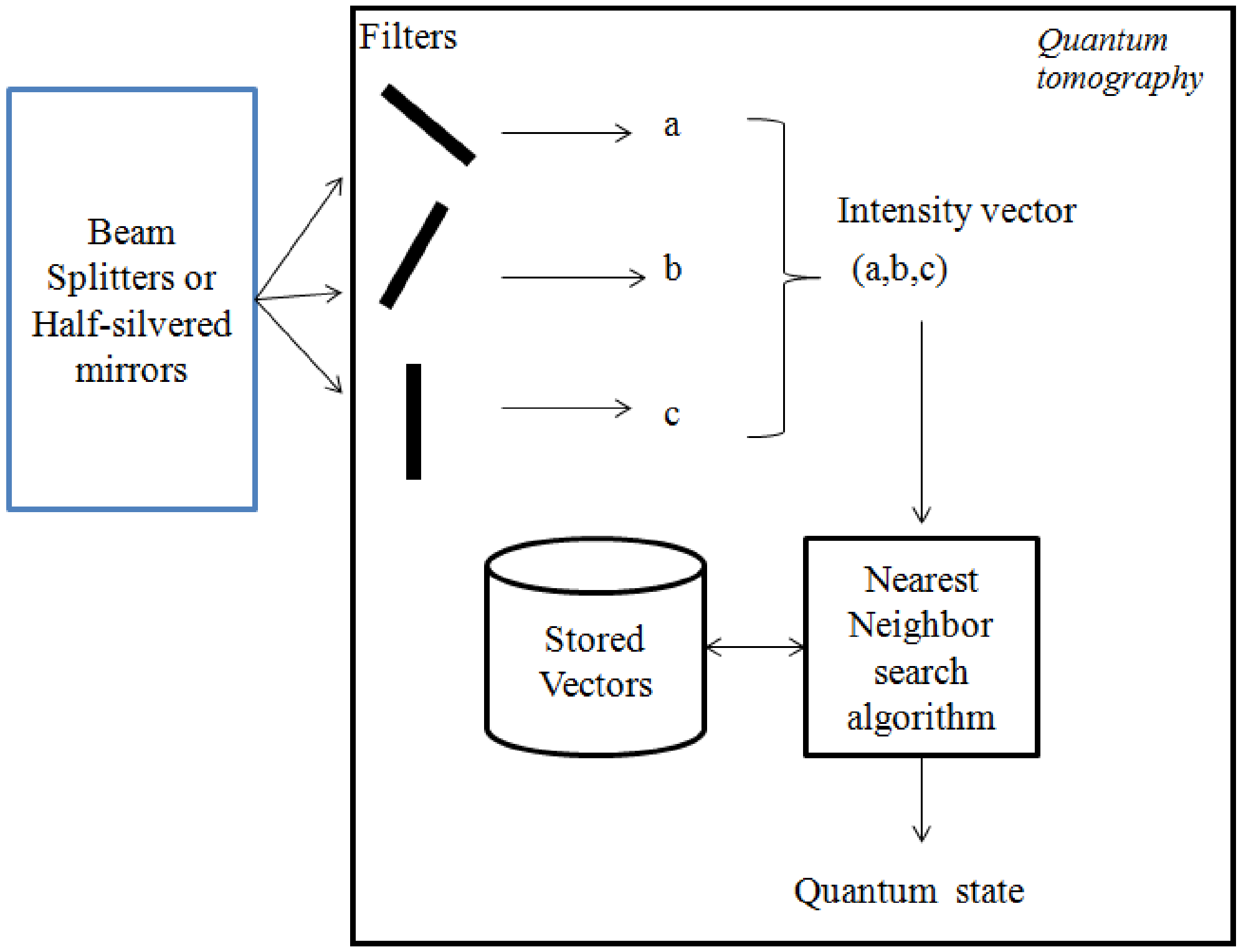

Figure 3 Quantum tomography process

We assume that Alice and Bob negotiate on a set of polarization angles that are used in the protocol. These set of polarization angles can be any of 4, 8, 16, 32, 64, 128, and so on, on the same plane. The possible polarization angles for each of these categories are shown in Table 1.

Table 1 Set of polarization angles used to represent bit 0 and 1

| Number of polarization angles | Possible polarization angles |
|---|---|
| 4 | $\{0^0, 45^0, 90^0, 135^0\}$ |
| 8 | $\{0^0, 22.5^0, 45^0, 67.5^0, 90^0, 112.5^0, 135^0, 157.5^0\}$ |
| 16 | $\{0^0, 11.25^0, 22.5^0, 33.75^0, 45^0, 56.25^0, 67.5^0, 78.75^0, 90^0, 101.25^0, 112.5^0, 123.75^0, 135^0, 146.25^0, 157.5^0, 168.75^0\}$ |

### 2.1 Filters, intensity vectors

A set of filters oriented at different angles are placed immediately after the beamsplitters. The stream of photons passes through the beamsplitter and splits in such a way that they pass through these filters. A filter allows photons with matching polarization angle to pass through them. Thus the output of the filter gives the intensity of the photons passed through them. These intensity values are put together to form intensity vectors. The intensity vector gives the number of photons that passed through each filter.

### 2.2 Stored vectors

The intensity vectors are computed by sending photons with each of the polarization angles from a given set. For example, consider a protocol where Alice and Bob negotiated to use a set of 8 polarization angles. In this case, a bunch of photons with polarization angle $0^0$ are sent through the filters and the resultant intensity vector is stored as $V_1$.

Similarly, a bunch of photons with polarization angle $22.5^0$ is sent through the filters and the intensity vector stored is $V_2$. Then, photons with polarization angle $45^0$, $67.5^0$, $90^0$, $112.5^0$, $135^0$, $157.5^0$ are sent through the filters resulting intensity vectors $V_3$, $V_4$, $V_5$, $V_6$, $V_7$ and $V_8$ respectively. So, in a scenario of 8 and 16 polarization angles, the intensity vectors stored at the receiver's site are shown in the Table 2 and Table 3 respectively.

### 2.3 Choice of filters

One of the interesting questions is the number of filters and the angles of filters that are required for successful detection of a quantum state. According to theory, a set of 4 polarization angles need 2 filters to obtain unique stored vectors. A set of 8, 16, 32, 64, 128 polarization angles require 3, 4, 5, 6 and 7 filters respectively. Thus, if the number of polarization angles is $2^m$, we need *m* number of filters to obtain unique stored vectors.

The angles at which these filters are oriented can be any of the angles from the set or they can deviate from them. But the optimal choice of filters for a set of 4 and 8 polarization angles is filters with angles which differ by $45^0$ and $30^0$ respectively. Similarly, the optimal choice for other sets can be found by our theoretical experiments.

### 2.4 Number of photons

Another interesting question is to find the number of photons needed to obtain unique integer stored intensity vectors. We did our experiments starting with a case where we sent one photon through each filter and then increasing the number until we get integer components for intensity vectors.

For illustration, consider a set of 4 polarization angles. We need 2 filters in this case and we begin our experiment by sending a total of 2 photons that is one photon through each filter. The resulting intensity vectors are unique but they are not integer values. Therefore in the next step we increase the number of photons sent through each filter by 1 until we get integer values of intensities at the filters. We obtain integer component vectors when 4 photons are sent through each filter that is a total of 8 photons which is $2^3$. Thus, we can say that we need $m^2$ photons at each filter and a total of $m^3$ photons for the tomography process. We find that in Tables 2 and 3, each of the intensity vectors is unique. This makes it easy to identify the polarization angle one the outputs of all the filters have been measured.

The results of the experiments are shown in Table 4 and we are lead to the following result:

> *If there are n number of polarization angles equal to $2^m$ in a fixed plane, we need m number of filters and $m^2$ number of photons through each filter. Calculations starting with the base case are provided in the following paragraphs.*

Table 2 Stored intensity vectors at receiver's site for a set of 8 polarization angles

| Polarization of the photons sent by Alice | Intensity at filter $0^0$ | Intensity at filter $22.5^0$ | Intensity at filter $45^0$ | Intensity vectors (Stored vectors) |
|---|---|---|---|---|
| $0^0$ | 3 | 3 | 2 | (3, 3, 2) |
| $22.5^0$ | 3 | 3 | 3 | (3, 3, 3) |
| $45^0$ | 1 | 3 | 3 | (1, 3, 3) |
| $67.5^0$ | 0 | 2 | 3 | (0, 2, 3) |
| $90^0$ | 0 | 0 | 1 | (0, 0, 1) |
| $112.5^0$ | 0 | 0 | 0 | (0, 0, 0) |
| $135^0$ | 2 | 0 | 0 | (2, 0, 0) |
| $157.5^0$ | 3 | 2 | 0 | (3, 2, 0) |

## 2.5 Nearest neighbor search algorithm

The nearest neighbor search algorithm helps to find the proximity between 2 given vectors. The Euclidean distance between two vectors, $(a_1, b_1, c_1)$ and $(a_2, b_2, c_2)$ is calculated as $\sqrt{(a_1 - a_2)^2 + (b_1 - b_2)^2 + (c_1 - c_2)^2}$ where $(a_1, b_1, c_1)$ is the stored vector and $(a_2, b_2, c_2)$ is the vector that is obtained when the photons sent by Alice are passed through filters.

## 2.6 Process of state detection

Consider an example where the vectors for each possible value of $\theta$ for a protocol which uses 8 polarization angles are stored at the Bob's site in the form of Table 2. When Alice sends a set of photons with polarization $\theta = \{0^0, 22.5^0, 45^0, 67.5^0, 90^0, 112.5^0, 135^0, 157.5^0\}$, at Bob's site, they are sent through the three filters to get the intensity vector, V. This V is compared with the existing vectors {(3 3 2), (3 3 3), (1 3 3), (0 2 3), (0 0 1), (0 0 0), (2 0 0), (3 2 0)}. When a match is found, the polarization is determined which gives the state of the photons. Even if there is slight deviation in the vector obtained, then the nearest neighbor search algorithm is applied to find the closest vector that matches. Once the matching vector is found, the corresponding angle determines the unknown state of the photons sent by Alice.

Table 3 Stored intensity vectors at receiver's site for a set of 16 polarization angles

| Polarization of the photons sent by Alice | Intensity at filter $22.5^0$ | Intensity at filter $45^0$ | Intensity at filter $67.5^0$ | Intensity at filter $90^0$ | Intensity vectors (stored vectors) |
|---|---|---|---|---|---|
| $0^0$ | 3 | 2 | 1 | 0 | (3,2,1,0) |
| $11.25^0$ | 4 | 3 | 1 | 0 | (4,3,1,0) |
| $22.5^0$ | 4 | 3 | 2 | 1 | (4,3,2,1) |
| $33.75^0$ | 4 | 4 | 3 | 1 | (4,4,3,1) |
| $45^0$ | 3 | 4 | 3 | 2 | (3,4,3,2) |
| $56.25^0$ | 3 | 4 | 4 | 3 | (3,4,4,3) |
| $67.5^0$ | 2 | 3 | 4 | 3 | (2,3,4,3) |
| $78.75^0$ | 1 | 3 | 4 | 4 | (1,3,4,4) |
| $90^0$ | 1 | 2 | 3 | 4 | (1,2,3,4) |
| $101.25^0$ | 0 | 1 | 3 | 4 | (0,1,3,4) |
| $112.5^0$ | 0 | 1 | 2 | 3 | (0,1,2,3) |
| $123.75^0$ | 0 | 0 | 1 | 3 | (0,0,1,3) |
| $135^0$ | 1 | 0 | 1 | 2 | (1,0,1,2) |
| $146.25^0$ | 1 | 0 | 0 | 1 | (1,0,0,1) |
| $157.5^0$ | 2 | 1 | 0 | 1 | (2,1,0,1) |
| $168.75^0$ | 3 | 1 | 0 | 0 | (3,1,0,0) |

Table 4 Number of photons and filters required in the tomography process

| Number of polarization angles | Number of filters | Total number of photons |
|---|---|---|
| 4 | 2 | 8 |
| 8 | 3 | 27 |
| 16 | 4 | 64 |
| 32 | 5 | 125 |
| 64 | 6 | 216 |
| 128 | 7 | 343 |

## 3. Conclusion

We considered the problem of practical quantum tomography using beamsplitters and filters and found that the number of photons required for easy implementation using nearest neighbor rule is $m^3$ if the number of polarization angles $2^m$. Clearly, such an implementation provides considerable protection in the use of multistage quantum cryptography protocols.